\documentclass[10pt,twocolumn]{article}

\usepackage[utf8]{inputenc}
\usepackage{amsmath}
\usepackage{amssymb}
\usepackage{amsthm}
\usepackage{units}
\usepackage{titlesec}
\usepackage{cite}
\usepackage{booktabs}
\usepackage{graphicx}
\usepackage[colorlinks=false]{hyperref}
\usepackage[format=plain,labelfont=it]{caption}
\usepackage[left=1.5cm,right=1.5cm,top=2cm,bottom=2cm]{geometry}
\usepackage{float}

\titleformat{\section}{\centering\normalfont\scshape}{\Roman{section}.}{5pt}{}
\titleformat{\subsection}{\normalfont\it}{\Alph{subsection}.}{5pt}{}
\titleformat{\subsubsection}{\normalfont\it}{\hspace{4mm}\arabic{subsubsection})}{5pt}{}

\newcommand\infoFootnote[1]{%
  \begingroup
  \renewcommand\thefootnote{}\footnote{#1}%
  \addtocounter{footnote}{-1}%
  \endgroup}

\newtheorem{thm}{Theorem}
\newtheorem{assum}{Assumption}
\newtheorem{remark}{Remark}

\newtheorem{prop}[thm]{Proposition}
\newtheorem{lem}[thm]{Lemma}
\newtheorem{exmp}{Example}

\usepackage{tikz}
\usetikzlibrary{arrows}
\usetikzlibrary{shapes}
\usetikzlibrary{decorations.pathmorphing}
\usetikzlibrary{decorations.pathreplacing}
\usetikzlibrary{decorations.shapes}
\usetikzlibrary{decorations.text}
\usetikzlibrary{positioning, shapes}

\tikzset{
block/.style = {draw, fill=white, rectangle, minimum height=3em, minimum width=4.5em},
tmp/.style  = {coordinate}, 
sum/.style= {draw, fill=white, circle, node distance=2cm},
input/.style = {coordinate},
output/.style= {coordinate}
pinstyle/.style = {pin edge={to-,thick,black}}
}

\newcommand{\R}{\mathbb{R}} 
\newcommand{\N}{\mathbb{N}}

\newcommand{\Xc}{\mathcal{X}}

\newcommand{\Fc}{\mathcal{F}} 
  
\newcommand{\Sc}{\mathcal{S}}
\newcommand{\Tc}{\mathcal{T}}
\newcommand{\Uc}{\mathcal{U}}

\newcommand{\Zc}{\mathcal{Z}}

\newcommand{\zb}{\boldsymbol{z}}

\newcommand{\xib}{\boldsymbol{\xi}}
\newcommand{\xb}{\boldsymbol{x}}

\newcommand{\bb}{\boldsymbol{b}}
\newcommand{\wb}{\boldsymbol{w}}
\newcommand{\fb}{\boldsymbol{f}}

\newcommand{\Qb}{\boldsymbol{Q}}

\newcommand{\cb}{\boldsymbol{c}}

\newcommand{\Pb}{\boldsymbol{P}}
\newcommand{\Ab}{\boldsymbol{A}}

\newcommand{\Tb}{\boldsymbol{T}}

\newcommand{\Ib}{\boldsymbol{I}}

\newcommand{\zerob}{\boldsymbol{0}}

\newcommand{\kappab}{\boldsymbol{\kappa}}
\newcommand{\alphab}{\boldsymbol{\alpha}}

\newcommand{\taub}{\boldsymbol{\tau}}

\newcommand{\interior}{\mathrm{int}}

\title{\vspace{-2mm}\bf 
Convex reformulations for a special class of nonlinear MPC problems}
\author{Manuel Klädtke and Moritz Schulze Darup\vspace{2mm}}
\date{}

\begin{document}

\maketitle

\textbf{\textit{Abstract}.} {\bf We show how the solution to NMPC problems for a special type of input-affine discrete-time systems can be obtained by reformulating the underlying non-convex optimal control problem in terms of a finite number of convex subproblems. The reformulation is facilitated by exact (input-state) linearization, which is shown to provide beneficial properties for the treated class of systems. We characterize possible types of the resulting convex subproblems and illustrate our approach with three numerical examples.}
\infoFootnote{M. Klädtke and M. Schulze Darup are with the \href{https://rcs.mb.tu-dortmund.de/}{Control and~Cyberphysical Systems Group}, Faculty of Mechanical Engineering, TU Dortmund University, Germany. E-mails:  \href{mailto:manuel.klaedtke@tu-dortmund.de}{\{manuel.klaedtke, moritz.schulzedarup\}@tu-dortmund.de}. \vspace{0.5mm}}
\infoFootnote{\hspace{-1.5mm}$^\ast$This paper is a \textbf{preprint} of a contribution to the 20th European Control Conference 2022. The DOI of the original paper is \href{https://doi.org/10.23919/ECC55457.2022.9838061}{10.23919/ECC55457.2022.9838061}.}

\section{Introduction}

Model predictive control (MPC) is a widely used and well-studied control scheme, in which a dynamical model is instrumental for predicting and optimizing future system behavior through suitable control actions. In particular, the classical realization with linear prediction models, polyhedral constraints, and quadratic cost functions is well understood. In fact, in this case that is usually referred to as linear MPC, we obtain (under mild conditions) a convex optimal control problem (OCP), which can be solved reliably, efficiently, and even explicitly \cite{Bemporad2002, Seron2003}. For nonlinear MPC (NMPC), convexity is usually lost due to nonlinear system dynamics \cite{Rawlings2017}, making it hard to reliably compute the global optimum. Therefore, the great watershed in optimization between convexity and non-convexity \cite{Rockafellar1993} is typically associated with linear and nonlinear MPC, respectively. Nevertheless, although quite rare, there exist setups (such as those in \cite{Mare2007,Azhmyakov2008,Lautenschlager2015,NMPC2021Exact}) for which convex NMPC formulations can be derived. However, these approaches usually build on quite restrictive assumptions.

In this paper, we contribute to the investigation of convex (or ``convexifiable'') NMPC by showing how NMPC problems for a special type of input-affine systems can be reformulated in terms of a finite number of convex subproblems. Similar to and building upon the approach \cite{NMPC2021Exact}, we make use of exact input-state linearization (see \cite{Isidori1995} for an overview) to derive our results. While this kind of system transformation often leads to a loss of convexity for cost functions and constraints (as, e.g., noted in \cite{Mayne2000}), we show that a certain decomposition of the constraints in combination with a tailored parametrization of the stage costs allows to obtain the desired convex reformulations. 

We organize the presentation of the novel approach as follows. We summarize basics on NMPC and exact linearization for discrete-time systems in Section~\ref{sec:basics}. In Section~\ref{sec:characterization}, we characterize a special type of input-affine systems and show how exact linearization can be applied to these systems in a way that is beneficial for solving an NMPC problem. We present our main result, i.e., the reformulation of special types of NMPC problems for the presented class of systems as a finite number of convex subproblems, in Section~\ref{sec:convexNMPC}. Finally, we apply our results to three different numerical examples in Section~\ref{sec:examples} and suggest directions for future research in Section~\ref{sec:summary}.

\section{Basics on NMPC and exact linearization}\label{sec:basics}

We initially summarize some basics on MPC and exact (input-state) linearization for the general class of discrete-time nonlinear single-input systems
\begin{equation}
\label{eq:nonlinearSystem}
\xb(k+1)=\fb(\xb(k),u(k))
\end{equation}
with state and input constraints
\begin{equation}
\label{eq:constraints}
\xb(k) \in \Xc \subset \R^n \qquad \text{and} \qquad u(k) \in \Uc\subset \R.
\end{equation}
We specify the special class of interest later in Section~\ref{sec:characterization}.

\subsection{Nonlinear model predictive control}

Applying MPC to a nonlinear system \eqref{eq:nonlinearSystem} with constraints~\eqref{eq:constraints} requires solving an optimal control problem (OCP) of the form
\begin{align}
\label{eq:NMPC}
V(\xb):= \!\!\!\! \min_{\substack{\hat{\xb}(0),\dots,\hat{\xb}(N),\\\hat{u}(0),\dots,\hat{u}(N-1)}} \!\!\!\!\!\!\!\!\!\!&\,\,\,\,\,\,\,\,\,\, \varphi(\hat{\xb}(N)) +\!\! \sum_{k=0}^{N-1} \ell(\hat{\xb}(k),\hat{u}(k))\!\!\!\!\!\!\!\!\!\!\!\!\!\span  \\
\nonumber
\text{s.t.} \qquad   \hat{\xb}(0)&=\xb, \\
\nonumber
\hat{\xb}(k+1)&=\fb(\hat{\xb}(k),\hat{u}(k)) &\forall k \in \{0,\dots,N\!-1\}, \\
 \nonumber
 (\hat{\xb}(k),\hat{u}(k)) & \in \Xc \times \Uc &\forall k \in \{0,\dots,N\!-1\},\\
 \nonumber
\hat{\xb}(N)& \in \Tc
\end{align}
in each time step $k$ for the current state $\xb=\xb(k)$.  Here, $N\in \N$ is the prediction horizon, $\ell: \R^n \times \R \to \R$ denotes the stage cost, and $\varphi: \R^n \to \R$ refers to the terminal cost. Further, the terminal set $\Tc \subseteq \Xc$ can be used to enforce stability guarantees \cite{Mayne2000}. Due to the nonlinear system dynamics that occur as equality constraints, the OCP is typically non-convex, even if convex cost functions and constraint sets are considered. Hence, evaluating the optimal control sequence $\hat u^\ast(0),\dots,\hat u^\ast(N-1)$ and, in particular, the control action $u(k)=\hat{u}^\ast(0)$, can be numerically hard.

\subsection{Exact linearization of discrete-time systems}

The exact (input-state) linearization of discrete-time systems~\eqref{eq:nonlinearSystem}, which from now on we will simply refer to as exact linearization, 
is less popular than its continuous-time counterpart. Hence, we briefly summarize the key ingredients and refer to, e.g., \cite{Soroush1992} for further details. In a nutshell, exact linearization requires finding a new set of coordinates $\xib := \taub(\xb)$ and a feedback $u := \Psi(\xb,v)$ with an artificial input~$v$, which transform \eqref{eq:nonlinearSystem} into an equivalent linear system
\begin{equation}
\label{eq:linearizedStateDynamics}
\xib(k+1)=\widetilde{\Ab} \xib(k)+ \tilde{\bb} \,v(k) .
\end{equation}
To achieve this, the relative degree $r$ with respect to an (artificial) output
\begin{equation}
    y(k) = h(\xb(k)) \label{eq:nonlinearOutput}
\end{equation}
plays a central role. Intuitively, $r$ can be seen as the minimum number of time steps after which the input $u$ affects $y$. More precisely, $r$ can be defined based on the recursion 
\begin{equation}
    h^{(i)}(\xb):=h^{(i-1)}(f(\xb,u)) \quad \text{with} \quad h^{(0)}(\xb):=h(\xb), \label{eq:recursion}
\end{equation}
where we omit the argument $u$, since all relevant $h^{(i)}(\xb)$ will not depend on $u$ by assumption. The relative degree of~\eqref{eq:nonlinearSystem} w.r.t.~\eqref{eq:nonlinearOutput} at $(\xb^\circ,u^\circ)$ is then defined as the $r\in\N$ that obeys
\begin{subequations}
\label{eq:conditionsRelDegree}
\begin{align}
\label{eq:conditionNonZero}
\!\!\!\frac{\partial}{\partial u} h^{(r-1)}(f(\xb^\circ,u^\circ)) &\neq 0 \,\,\, \text{and}\\ 
\label{eq:conditionZero}
\frac{\partial}{\partial u} h^{(i)}(\xb) &= 0 \,\,\, \text{for every} \,\, i \in \{0,\dots,r-1\}\!\!
\end{align}
\end{subequations}
and every $(\xb,u)$ in a neighborhood of $(\xb^\circ,u^\circ)$, where condition \eqref{eq:conditionZero} justifies the notation in \eqref{eq:recursion}.
For the special case $r = n$, the state transformation
\begin{equation}
    \label{eq:nonlinearTransformation}
    \taub(\xb):=\begin{pmatrix}
    h^{0}(\xb) &
    \hdots &
    h^{n-1}(\xb)
    \end{pmatrix}^\top
\end{equation}
yields the transformed state space system
\begin{align}
\nonumber
\xib(k+1)&=\taub(\fb(\xb(k),u(k)))\\
&= \begin{pmatrix}
\xib_2(k) \\
\vdots \\
\xib_n(k) \\
h^{(n-1)}(\fb(\xb(k),u(k)))\label{eq:zStep}
\end{pmatrix}. 
\end{align}
Regarding the $n$-th entry of~\eqref{eq:zStep}, we note that \eqref{eq:conditionNonZero} implies the existence of a function $\Psi_y: \R^n \times \R \rightarrow \R$ such that
\begin{equation}
\label{eq:PsiImplicitlyDefined}
h^{(n-1)}(\fb(\xb,\Psi_y(\xb,y)))= y
\end{equation}
holds (at least locally) by the implicit function theorem. Now, let us consider the feedback $u(k)=\Psi(\xb(k),v(k))$ with
\begin{equation}
\label{eq:Psi}
\Psi(\xb,v):=\Psi_y\left(\xb,b_0 v - \sum_{i=0}^{n-1} a_i h^{(i)}(\xb)\right)
\end{equation}
for some arbitrary coefficient $b_0 \neq 0$ and ${a_0,\dots,a_{n-1} \in \R}$.
Then, we obtain
\begin{equation}
\nonumber
h^{(n-1)}(\fb(\xb(k),\Psi(\xb(k),v(k))) =b_0 v(k) - \sum_{i=0}^{n-1} a_i  \xib_{i+1}(k),
\end{equation}
according to~\eqref{eq:PsiImplicitlyDefined}.
Together with \eqref{eq:zStep}, this finally leads to the parameters
$$
\widetilde{\Ab} \!:=\! \begin{pmatrix}
0 & 1 & 0 & \cdots & 0 \\
0 & 0 & 1 & & 0 \\
\vdots & \vdots & & \ddots  \\
0 & 0 & 0 & & 1 \\
-a_0 & -a_1 & -a_2 & \cdots & -a_{n-1}
\end{pmatrix}\!, \,\, \tilde{\bb}:=\!\begin{pmatrix}
0 \\
0 \\
\vdots \\
0 \\
b_0
\end{pmatrix}\! 
$$
for the exactly linearized system \eqref{eq:linearizedStateDynamics}.

\section{System specification and exact linearization}
\label{sec:characterization}

We begin our analysis with a characterization of the special system class for which our methodology will apply.

\begin{assum}\label{assum:characterization}
We consider systems~\eqref{eq:nonlinearSystem} with
\begin{equation}
\label{eq:specialDynamics}
\fb(\xb,u):=\Ab \xb + g(\xb)  \bb \, u,
\end{equation}
where $\Ab \in \R^{n\times n}$, $\bb \in \R^n$, and $g:\R^n \rightarrow \R$. We assume that  the pair $(\Ab,\bb)$ is controllable and that $g(\zerob)\neq 0$. We further assume that $\Xc$ can be decomposed into $s \in \N$ convex subsets $\Xc_i$, i.e.,
\begin{equation}
\label{eq:XDecompositon}
\Xc := \bigcup_{i=1}^s \Xc_i
\end{equation}
with the following properties: For every $i \in \{1,\dots,s\}$, we either have
\begin{subequations}
\label{eq:concaveConvex}
\begin{align}
  \!\!\!g(\xb)&\geq 0, \,\,\, g(\eta \xb + (1-\eta)\hat{\xb}) \geq \eta  g(\xb) \!+\! (1-\eta)g(\hat{\xb})\!\!\!\\
  \nonumber
  \text{or} & \\
  \!\!\!g(\xb)&\leq 0, \,\,\, g(\eta \xb + (1-\eta)\hat{\xb}) \leq \eta  g(\xb) \!+\! (1-\eta)g(\hat{\xb})\!\!\!
\end{align}
\end{subequations}
for every $\xb,\hat{\xb} \in \Xc_i$ and every $\eta \in (0,1)$. Finally, we assume that $\zerob \in \interior(\Xc_1)$ and that $\Uc:=[\underline{u},\overline{u}]$ with $\underline{u}< 0 < \overline{u}$.
\end{assum}

Without doubt, these assumptions are restrictive. Nevertheless, they can be easily interpreted and they include interesting system classes. For instance, \eqref{eq:specialDynamics} describes input-affine systems with a linear drift and ``linearly dependent'' input-nonlinearities. Moreover, \eqref{eq:concaveConvex} states that $g$ should either be non-negative and concave or non-positive and convex on each subset~$\Xc_i$. Both assumptions are, e.g., satisfied by the special class of bilinear systems in \cite{NMPC2021Exact}. In fact, we there have  $g(\xb)=\wb^\top \xb + d$ with the specifications $\wb \neq \zerob$ and $d\neq 0$ (in order to exclude trivial linear systems and $g(\zerob)=0$). In addition, $\Xc$ is assumed to be convex with $\zerob \in \interior(\Xc)$ in \cite{NMPC2021Exact}. Hence, the sets 
\begin{align*}
      \Xc_1&:= \{ \xb \in \Xc \,|\, d \,g(\xb) \geq 0 \} \quad \text{and} \\
      \Xc_2&:= \{ \xb \in \Xc \,|\, d \,g(\xb) \leq 0 \}  
\end{align*}
are both convex and $g$ satisfies~\eqref{eq:concaveConvex} (since affine functions are obviously both convex and concave and since the sign of $g(\xb)$ is used to specify $\Xc_1$ and $\Xc_2$). Finally, $\zerob \in \interior(\Xc_1)$ is satisfied by construction (since $d\,g(\zerob)=d^2>0$).
Now, Assumption~\ref{assum:characterization} can also be met for piecewise affine $g$, which significanly extends the former case. In fact, assuming that the affine segments are defined over convex domains, we can easily identify convex sets $\Xc_i$ on which~\eqref{eq:concaveConvex} holds.
Finally, Assumption~\ref{assum:characterization} can be satisfied for functions $g$ that reflect a composition of a real function consisting of either non-negative and concave or non-positive and convex segments with an affine mapping.
For an example, such a construction underlies the function $g(\xb)=4\cos\left(\nicefrac{3\pi}{8}(\xb_1-\xb_2)\right)$ that will be investigated in more detail further below.

Now, for the exact linearization of a system~\eqref{eq:nonlinearSystem} with the specification~\eqref{eq:specialDynamics}, we consider a linear output function $h(\xb) := \cb^\top\xb$, where $\cb$ is chosen such that 
\begin{subequations}
\label{eq:cAbConditions}
\begin{align}
    \!\!\cb^\top \!\Ab^{n-1} \bb &\neq 0 \,\,\, \text{and} \\
    \cb^\top \Ab^{i} \bb &= 0 \,\,\, \text{for every}\,\,i \in \{0,\dots,n-2\}.
\end{align}
\end{subequations}
We note that such a choice is always possible since $(\Ab,\bb)$ is controllable according to Assumption~\ref{assum:characterization}.
Based on~\eqref{eq:cAbConditions}, it is easy to verify that \eqref{eq:conditionsRelDegree} holds for 
$$
\xb^\circ \in \Xc^\circ:=\{ \xb \in \R^n \,|\, g(\xb) \neq \zerob \}.
$$
Hence, the system has relative degree $r = n$ (w.r.t. the chosen output) almost everywhere.
Further, due to the linear output and the linear drift, the transformation \eqref{eq:nonlinearTransformation} is likewise linear and results in
\begin{equation}
\label{eq:linearTransformation}
\taub(\xb)= \Tb \xb \quad \text{with} \quad \Tb:= \begin{pmatrix}
\cb^\top \!\Ab^0 \\
\vdots \\
\cb^\top \!\Ab^{n-1} 
\end{pmatrix}.
\end{equation}
Moreover, \eqref{eq:PsiImplicitlyDefined} in combination with~\eqref{eq:Psi} yields
\begin{equation}
\label{eq:uImplicit}
    \cb^\top \Ab^{n-1} \left(\Ab \xb + g(\xb)\, \Psi(\xb,v)\right)=b_0 v-\!\!\sum_{i=0}^{n-1} a_i \cb^\top\Ab^i \xb. 
\end{equation}
Hence, the linearizing feedback $\Psi(\xb,v)$ can be specified as
\begin{equation}
\label{eq:PsiFrac}
 \Psi(\xb,v):=\frac{b_0 v-\alphab^\top \xb}{\beta g(\xb)}
\end{equation}
with
$$
 \alphab^\top := \cb^\top\!\left( \Ab^n + \sum_{i=0}^{n-1} a_i \Ab^i\right)  \quad \text{and} \quad \beta := \cb^\top \Ab^{n-1} \bb,
$$
where we note that the fraction in~\eqref{eq:PsiFrac} is well-defined for every $\xb \in \Xc^\circ$.

Since \eqref{eq:linearTransformation} is a linear transformation, it can be reverted while maintaining linear dynamics, resulting in
\begin{equation}
\label{eq:linSysOriginalStates}
\xb(k+1)=\widehat{\Ab} \xb(k)+\hat{\bb} v(k)
\end{equation}
with $\widehat{\Ab}:=\Tb^{-1} \widetilde{\Ab} \Tb$ and $\hat{\bb}:=\Tb^{-1} \tilde{\bb}$. This traces the coordinates of the exactly linearized system back to the original state-space, which proves helpful regarding the use of exact linearization in conjunction with MPC, since both the formulation of state constraints and cost functions can be based on the original states.

\section{Convex reformulations via decomposition}\label{sec:convexNMPC}

The central idea now is to reformulate the original OCP~\eqref{eq:NMPC} using~\eqref{eq:linSysOriginalStates} and to exploit the resulting structure. To this end, we first address reformulations of the constraints and cost functions  in~\eqref{eq:NMPC}. Fortunately, since~\eqref{eq:linSysOriginalStates} builds on the original states, the terminal cost $\varphi$ and the terminal set $\Tc$ will be the same.
In contrast, the stage cost and the input (and state) constraints have to be reformulated in order to incorporate the novel input $v$ (instead of $u$). Taking the linearizing feedback  $u=\Psi(\xb,v)$ into account, the reformulated stage cost and constraints will, in principle, follow from
$\ell(\xb, \Psi(\xb,v))$ and \vspace{2mm}
\begin{equation}
    \label{eq:illDefiniedZ}
    \left\{ \left.\begin{pmatrix}
\xb \\
v 
\end{pmatrix} \in \R^{n+1} \right|\, \xb \in \Xc ,\, \Psi(\xb,v) \in \Uc \right\}. \vspace{2mm}
\end{equation}
However, both expressions are well-defined only for $\xb \in \Xc^\circ$. As a remedy, we first introduce the novel OCP \vspace{2mm}
\begin{align}
\label{eq:MPC}
\!\!V(\xb):= \!\!\!\! \min_{\substack{\hat{\xb}(0),\dots,\hat{\xb}(N),\\\hat{v}(0),\dots,\hat{v}(N-1)}} \!\!\!\!\!\!\!\!\!\!\!\!\!&\,\,\,\,\,\,\,\,\,\,  \varphi(\hat\xb(N)) +\!\! \sum_{k=0}^{N-1} \! \hat\ell(\hat{\xb}(k),\hat{v}(k))\!\!\!\!\!\!\!\!\span   \\
\nonumber
\text{s.t.} \qquad   \hat{\xb}(0)&=\xb, \\
\nonumber
\hat{\xb}(k+1)&=\widehat{\Ab} \hat{\xb}(k)+\hat{\bb}\hat{v}(k) &\forall k \in \{0,\dots,N\!-1\}, \\
 \nonumber
 \begin{pmatrix}
 \hat{\xb}(k)\\
 \hat{v}(k)
 \end{pmatrix}\! & \in \Zc &\forall k \in \{0,\dots,N\!-1\},\\
 \nonumber
\hat{\xb}(N)& \in \Tc \vspace{4mm}
\end{align}
and discuss suitable choices for the novel stage cost $\hat{\ell}$ and the novel constraints $\Zc$ in the following. \vspace{2mm}

\subsection{Suitable constraints and their convex decomposition}

In order to fix the ill-definition of \eqref{eq:illDefiniedZ}, we first note that $u=\Psi(\xb,v)$ implies \vspace{2mm}
\begin{equation}
    \label{eq:vDependingOnU}
v=\frac{\beta g(\xb) u + \alphab^\top \xb}{b_0}.\vspace{2mm}
\end{equation}
Remarkably, this relation (which could also be derived from \eqref{eq:uImplicit}) is well-defined for every $\xb \in \R^n$. Now, whenever $\xb \notin \Xc^\circ$, we have $g(\xb)=0$ by definition. Hence, \eqref{eq:vDependingOnU} suggest to choose $v=\alphab^\top \xb/b_0$ in these cases.
Based on this observation and inspired by~\eqref{eq:illDefiniedZ}, we define $\Zc$ as \vspace{4mm}
\begin{align}
\nonumber
   \Zc &:=  \left\{ \left.\begin{pmatrix}
\xb \\
v 
\end{pmatrix} \in \R^{n+1} \right|\, \xb \in \Xc \cap \Xc^\circ ,\, \Psi(\xb,v) \in \Uc \right\} \\ 
\label{eq:setZ}
&\quad \cup \left\{ \left.\begin{pmatrix}
\xb \\
v 
\end{pmatrix} \in \R^{n+1} \right|\, \xb \in \Xc \setminus \Xc^\circ ,\, v = \frac{\alphab^\top \xb}{b_0} \right\}. \vspace{4mm}
\end{align}
One can then easily prove the following statements, which confirm the equivalence of the original and the novel constraints.

\begin{lem}\label{lem:equivalent_constraints}
Consider any $\xb\in\Xc$ and $u \in \Uc$ and define $v$ as in~\eqref{eq:vDependingOnU}. Then,
\begin{equation}
\label{eq:xvInZ}
     \begin{pmatrix}
\xb \\ 
v
\end{pmatrix}\in \Zc.
\end{equation}
Analogously, consider any $\xb \in \Xc$ and $v \in \R$ satisfying~\eqref{eq:xvInZ}. Then, 
$$
u := \left\{ \begin{array}{ll}
\Psi(\xb,v) & \text{if} \,\,\, \xb \in \Xc^\circ, \\
0 & \text{otherwise}
\end{array}\right.
$$
is such that~\eqref{eq:vDependingOnU} holds and $u \in \Uc$.
\end{lem}

It is easy to see that the specified $\Zc$ is usually non-convex. Nevertheless, $\Zc$ can be decomposed into finitely many convex subsets as specified in the following theorem.

\begin{thm}
Let Assumption~\ref{assum:characterization} hold. Then, $\Zc$ as in~\eqref{eq:setZ} can be decomposed into $s\in\N$ convex subsets 
\begin{equation}
    \label{eq:Zi}
    \!\Zc_i:=\left\{ \!\left.\begin{pmatrix}
\xb \\
v 
\end{pmatrix} \!\in \R^{n+1}\, \right| \xb \in \Xc_i ,\,   b_0 v - \alphab^\top \xb \in \Uc_i(\xb) \right\}\!
\end{equation}
i.e., $\Zc = \bigcup_{i=1}^s \Zc_i$, where 
$$
\Uc_i(\xb):=\left\{ \begin{array}{ll}
\,\![\beta g(\xb) \underline{u}, \beta g(\xb) \overline{u}]  & \text{if} \,\, \beta g(\xb)\geq 0 \,\,\forall \xb \in \Xc_i. \\[1mm]
\,\![\beta g(\xb) \overline{u}, \beta g(\xb) \underline{u}]  & \text{otherwise}. 
\end{array}
\right.
$$
\end{thm}

\begin{proof}
We first prove that 
\begin{align}
\nonumber
   \Zc_i &=  \left\{ \left.\begin{pmatrix}
\xb \\
v 
\end{pmatrix} \in \R^{n+1} \right|\, \xb \in \Xc_i \cap \Xc^\circ ,\, \Psi(\xb,v) \in \Uc \right\} \\ 
\label{eq:ZiEquivalence}
&\quad \cup \left\{ \left.\begin{pmatrix}
\xb \\
v 
\end{pmatrix} \in \R^{n+1} \right|\, \xb \in \Xc_i \setminus \Xc^\circ ,\, v = \frac{\alphab^\top \xb}{b_0} \right\}. 
\end{align}
To this end, we note that $\xb \in \Xc_i \cap \Xc^\circ$ implies
$$
\Psi(\xb,v) \in \Uc  \quad \Longleftrightarrow \quad b_0 v - \alphab^\top \xb \in \Uc_i(\xb)
$$
 and that  ${\xb \in \Xc_i \setminus \Xc^\circ}$ implies $\Uc_i(\xb)=[0,0]$.
 
Clearly, the latter implies $v=\alphab^\top \xb/b_0$, which already proves~\eqref{eq:ZiEquivalence}. In combination with~\eqref{eq:XDecompositon} and~\eqref{eq:setZ}, this immediately leads to $\Zc = \bigcup_{i=1}^s \Zc_i$. It remains to prove convexity of $\Zc_i$. To this end, we note that $\Zc_i$ as in~\eqref{eq:Zi} can  be equivalently characterized by the conditions
 \begin{subequations}
 \label{eq:setOfConditions}
 \begin{align}
   \xb &\in \Xc_i, \\
   \label{eq:firstInequality}
     \beta g(\xb) \underline{u} - b_0 v + \alphab^\top \xb &\leq 0, \\
     \label{eq:secondInequality}
     -\beta g(\xb)\overline{u} + b_0 v - \alphab^\top \xb &\leq 0
 \end{align}
 \end{subequations}
whenever $\beta g(\xb)\geq 0$ for every $\xb \in \Xc_i$. Now, according to Assumption~\ref{assum:characterization}, we either have $g(\xb)\geq 0$ and concavity of $g$ or $g(\xb)\leq 0$ and convexity of $g$ on $\Xc_i$. Further, $\beta g(\xb)\geq 0$ appears for either $\beta>0$ and $g(\xb)\geq 0$ or $\beta<0$ and $g(\xb)\leq 0$. In either case, the left-hand sides in~\eqref{eq:firstInequality} and \eqref{eq:secondInequality} describe convex functions in $\xb$ and $v$ and, hence, the associated constraints are convex. For instance, $\beta>0$ requires $g(\xb)\geq 0$ and consequently concavity of $g$, which implies that both $\beta g(\xb) \underline{u}$ and $-\beta g(\xb)\overline{u}$ are convex due to $\underline{u}<0<\overline{u}$. Since also $\Xc_i$ is convex by assumption, all conditions~\eqref{eq:setOfConditions} are convex, which implies convexity of $\Zc_i$. Finally, the case $\beta g(\xb)\leq 0$ can be proven analogously.
\end{proof}

\subsection{Choosing the cost functions and the terminal set}\label{subsec:cost_and_T}

As previously mentioned, the expression of the exactly linearized system dynamics \eqref{eq:linSysOriginalStates} in terms of the original state $\xb$ allows for a very natural choice of the terminal ingredients $\varphi(\xb)$ and $\Tc$. However, we will first discuss the more complex stage cost $\hat\ell(\xb,v)$ in more detail and examine both its design as well as the consequences for the stage cost $\ell(\xb,u)$ of the original OCP~\eqref{eq:NMPC}.

We choose $\hat\ell(\xb,v)$ to be convex on every $\Zc_i$, which assures convexity of the subproblems later introduced in \ref{subsec:scenarios}, and positive definite, which allows for the following proposition.
\begin{prop}\label{prop:posdef}
    Let Assumption 1 hold and let $\hat\ell(\xb,v)$ be positive definite. Then $\ell(\xb,u)$ is positive definite.
\end{prop}
\begin{proof}
    We have  
    \begin{align*}
        \ell(\xb, u) = \hat\ell\left(\xb,\frac{\beta g(\xb)u+\alphab^\top\xb}{b_0} \right) 
    \end{align*}
    from \eqref{eq:vDependingOnU} and  \vspace{-1mm}
    \begin{align*}
        \hat\ell(\xb,v) &=  0 \quad &&\text{for}\, (\xb,v) = (\zerob,0) \\
        \hat\ell(\xb,v ) &> 0 \quad &&\text{otherwise}  
    \end{align*}
    since $\hat\ell(\xb, v)$ is positive definite. But since $g(\xb = \zerob) \neq 0$ as per Assumption~\ref{assum:characterization}, the origin $(\xb, u) = (\zerob,0)$ is only mapped to the origin $(\xb, v) = (\zerob,0)$ and vice versa, which implies 
    \begin{align*}
        \ell(\xb,u) &=  0 \quad &&\text{for}\, (\xb,u) = (\zerob,0) \\
       \ell(\xb,u ) &> 0 \quad &&\text{otherwise} 
    \end{align*}
    and thus positive definiteness of $\ell(\xb,u)$.
\end{proof}
Designing the stage cost $\hat\ell(\xb,v)$ with respect to the artificial input $v$, it is important to examine its effect on the actual stage cost $\ell(\xb,u)$. For simplicity, consider $\beta = b_0 = 1$ and $\alphab = \zerob$, leading to $\ell(\xb, u) = \hat\ell(\xb, v = g(\xb) u)$, allowing for a straightforward interpretation. A stage cost like this does not penalize the actual input $u$, but rather its effect on the system given the current state $\xb$. In particular, this means that for states $\xb$, where $g(\xb) \approx 0$ and thus a given actual input $u$ barely affects the system, it also essentially becomes free of cost, which is certainly an unwanted feature. We note, however, that due to adhering to the input constraint $u \in \Uc$, the presented control approach does not allow the input $u$ to become too aggressive even for states $\xb$ where it becomes very cheap. Furthermore, although in a different setting, arguments similar to those in in  \cite[Sect. 4]{Freeman1995} suggest that $\ell(\xb,u)$ can represent sensible stage costs.

For the construction of the terminal set $\Tc$, we disregard every convex subset except $\Zc_1$ since, per Assumption 1, we have $\zerob \in \interior{(\Xc_1)}$ as well as $0\in \Uc$ and can thus conclude $\zerob \in \Zc_1$ from \eqref{eq:ZiEquivalence}. 
We then choose $v = \kappa(\xb)$ as a stabilizing  control law for the exactly linearized system and compute a convex terminal set $\Tc$ and convex terminal cost function $\varphi(\xb)$ such that the following axioms are satisfied 
\begin{subequations}
\begin{align}
    \begin{pmatrix}
        \xb^\top & \kappa(\xb)    
    \end{pmatrix}^\top
    &\in \Zc_1, \\ 
    \widehat\Ab\xb+\hat\bb\kappa(\xb) &\in \Tc, \\ 
    \varphi\!\left(\!\widehat\Ab\xb\!+\!\hat\bb\kappa(\xb)\!\right)\!-\!\varphi(\xb)\!+\!\hat\ell\left(\xb,\kappa(\xb)\right) &\leq 0 
\end{align}\label{eq:stability_axioms}%
\end{subequations}
for every $\xb \in \Tc$, which ensures stability of the closed-loop system in accordance with \cite{Mayne2000}.

\subsection{Scenario-based evaluation} \label{subsec:scenarios}

\begin{figure}
    \centering
    \resizebox{0.95\linewidth}{!}{\begin{tikzpicture}[level/.style={sibling distance=18mm/#1}]
\node  (z){}
    child {node  {$\Zc_1$}
        child {node  {$\Zc_1$}
            child {node (a)  {$\Zc_1$}}
            child {node   {$\cdots$}}
            child {node (b)   {$\Zc_s$}}
        }
        child {node   {$\cdots$}}
        child {node   {$\Zc_s$}}
    }
    child {node   {$\cdots$}}
    child {node   {$\Zc_s$}
        child {node  {$\Zc_1$}}
        child {node   {$\cdots$}}
        child {node   {$\Zc_s$}
            child {node (c)  {$\Zc_1$}}
            child {node   {$\cdots$}}
            child {node (d)   {$\Zc_s$}
                child [grow=right] {node (q) {$k = N-1$} edge from parent[draw=none]
                child [grow=up] {node (r) {$\vdots$} edge from parent[draw=none]
                child [grow=up] {node (t) {$k = 0$} edge from parent[draw=none]
                  child [grow=up] {node (u) {} edge from parent[draw=none]}
                }
              }
            }
            }
        }
    };
\path (b) -- (c) node (x) [midway] {$\cdots$}
  child [grow=down] {
    node (y) {}
    edge from parent[draw=none]
  };
 \draw[decoration={brace,mirror,raise=10pt},decorate]
  (a.west) -- node[below=15pt] {$s^N$} (d.east);
\end{tikzpicture}}
    \vspace{-15mm}
    \caption{A tree visualizing the $s^N$ different constraint scenarios, where in each time step $k$ one of the constraints $z(k)\in\Zc_i$ must hold.}\vspace{-6mm}
    \label{fig:scenario_tree}
\end{figure}
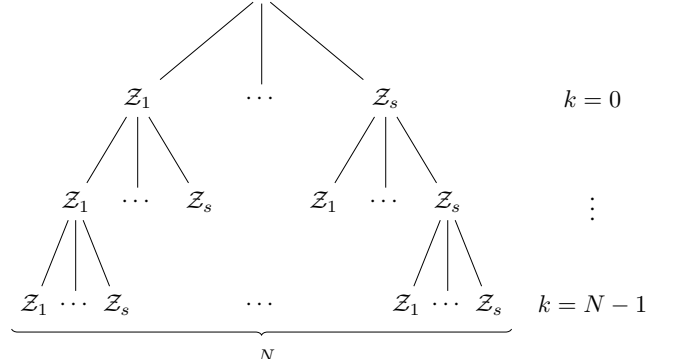

Having the exactly linearized model \eqref{eq:linSysOriginalStates}, convex costs $\hat\ell(\xb,v)$, $\varphi(\xb)$, a convex terminal set $\Tc$ and convex constraint subsets $\Zc_i$ is still not sufficient for \eqref{eq:MPC} to be a convex OCP, since the union of constraints $\Zc$ is typically non-convex. However, the condition
$$
    \begin{pmatrix}
    \hat{\xb}(k)\\
    \hat{v}(k)
    \end{pmatrix}\!  \in \Zc =
     \bigcup_{i=1}^s \Zc_i
$$
for each prediction step $k \in\{0, ..., N-1\}$ is equivalent to
$$
    \begin{pmatrix}
    \hat{\xb}(k)\\
    \hat{v}(k)
    \end{pmatrix}\!  \in \Zc_1 \quad
    \text{or} \quad
    ... \quad
    \text{or} \quad \begin{pmatrix}
    \hat{\xb}(k)\\
    \hat{v}(k)
    \end{pmatrix}\!  \in \Zc_s, 
$$
resulting in $s^N$ different scenarios, visualized as a tree in Figure \ref{fig:scenario_tree}. The sequence of convex constraints for a scenario $j \in \{1, ..., s^N\}$ can be compactly represented as 
$$
    \begin{pmatrix}
    \hat{\xb}(k)\\
    \hat{v}(k)
    \end{pmatrix}\!  \in \Zc_{\varepsilon_k(j)}
$$
with unique coefficients $\varepsilon_0(j), ..., \varepsilon_{N-1}(j) \in \{1, ..., s\}$ and
$$
    j = 1+\sum_{k=0}^{N-1} (\varepsilon_{k}(j)-1) s^k.
$$
Therefore, the solution to \eqref{eq:MPC} can be computed according to
$$
    V(\xb) = V^{(j^\ast)}(\xb)
$$
by finding the optimal constraint scenario
$$
    j^\ast := \arg \min_{j}\, V^{(j)}(\xb).
$$
Thus, the exact solution to the nonconvex OCP can be obtained by solving a finite number of convex subproblems.
\begin{remark}
    This strategy of solving the OCP can be seen as related to similar strategies for PWA systems, such as the one shown in  \cite{Mayne2003OptimalCO}. There, the OCP is evaluated for different switching sequences, which correspond to the constraint scenarios $j$ considered here. We note that, similar to the dynamics in PWA systems, we can also consider the case where the parameters $\Ab$ and $\bb$ may change for different subsets $\Xc_i$ without too much additional computational cost.
\end{remark}

Depending on $g(\xb)$ and the chosen cost functions $\hat\ell(\xb, v)$, $\varphi(\xb)$, these subproblems can become certain specific types of optimization problems. Having a (piecewise) affine $g(\xb)$ and quadratic costs result in QPs as shown in \cite{NMPC2021Exact}, while a (piecewise) quadratic $g(\xb)$ that fulfills Assumption 1 and quadratic costs result in convex quadratically constrained quadratic programs (QCQPs). Other types of functions typically result in more general nonlinear programs, which are, however, still convex.

\subsection{Considerations for computational complexity}

Having to solve $s^N$ subproblems is certainly computationally expensive, especially for large $N$. However, there are strategies that significantly reduce the effort required for online computation by preparing the solution offline. Since each constraint scenario $j$ corresponds to certain trajectories in the state-space, a lot of scenarios might not be feasible for any initial state $\hat\xb(0) = \xb$. These scenarios can be identified offline by treating the subproblems as multiparametric programs with parameter $\xb$, which allows to fully disregard infeasible scenarios during online computation.

For the identification of infeasible scenarios, it is beneficial to use a recursive approach based on the observation that having an infeasible scenario 
$$
    \tilde{j} = 1+\sum_{k=0}^{\tilde{N}-1} (\tilde{\varepsilon}_{k}(\tilde{j})-1) s^k.
$$
with prediction horizon $\tilde{N}$ implies infeasibility for all scenarios $j$ with horizon $N = \tilde{N}+1$, coefficients $\varepsilon_{k}({j}) = \tilde{\varepsilon}_{k-1}(\tilde{j}), k \in \{1,..., N-1\}$ and arbitrary $\varepsilon_{0}({j}) \in \{1, ..., s\}$. Thus, starting from $\tilde{N} = 1$ and gradually removing infeasible scenarios, we obtain many additional removed scenarios that do not need to be individually checked for feasibility. 

Furthermore, given an initial state $\hat\xb(0) = \xb$, all constraint scenarios $j$ where $\varepsilon_{0}({j}) = i$ and $\xb \notin \Xc_i$ can be neglected without checking feasibility, since due to \eqref{eq:ZiEquivalence}, we have
$$
    \xb \notin \Xc_i \quad \implies \quad \zb \notin \Zc_i
$$
for any $v$.

\section{Numerical examples}\label{sec:examples}

We present three numerical examples of a quadratic, a sinusoidal, and a PWA function $g(\xb)$ leading to convex subproblems that take the form of a QCQP, a general convex nonlinear program, and a QP, respectively. Similar to \cite{NMPC2021Exact}, we consider a two-dimensional example with parameters
\vspace{-2mm}
$$
    \Ab:=\begin{pmatrix}
        1.0 &\, 0.1 \\
        0.1 &\, 1.0 
    \end{pmatrix} \quad \text{and} \quad \bb:=\begin{pmatrix}
        0.01 \\
        0.05
    \end{pmatrix}\vspace{-2mm}
$$
as well as constraints
\vspace{-2mm}
$$
\Xc:=\{ \xb \in \R^2 \,|\, -2 \leq \xb_i \leq 2 \} \quad \text{and} \quad \Uc:=[-2,2] \vspace{-2mm}
$$
for each of the three examples to obtain comparable results. Regarding the exact linearization we also choose  $b_0 = 0.1$ as well as $\cb^\top = \begin{pmatrix} 5 & -1 
\end{pmatrix}$, which results in $\beta = \cb^\top \Ab \bb = 0.024$, and $\alphab = \zerob$ which requires $a_0 = 0.99$ and $a_1 = -2$. These choices result in the parameters
$$
    \widehat\Ab = \Ab \quad \text{and} \quad \hat\bb = 
    \begin{pmatrix}
    0.0417 \\ 0.2083
    \end{pmatrix}
$$
of the exact linearized system for all three examples. Remarkably, the shown procedure does not depend upon $g(\xb)$ and by our choice all three examples have the same exactly linearized dynamics but will eventually differ regarding their transformed constraints $\Zc$ and the effect of the stage cost function $\hat\ell(\xb,v)$ on $\ell(\xb,u)$. Regarding the cost functions, we choose a quadratic stage cost
$$
    \hat\ell(\xb,v) = \xb^\top \Qb \xb+\rho v^2
$$
with 
$$
    \Qb = \begin{pmatrix}
        0.05 & 0.00 \\ 0.00 & 0.05 
    \end{pmatrix} \quad \text{and} \quad \rho = \frac{0.1 b_0^2}{\beta^2}= 1.736,
$$ 
as well as a quadratic terminal cost $\varphi(\xb) = \xb^\top \Pb \xb$, where $\Pb$ is the solution of the Riccati equation
\begin{equation}
\label{eq:DARE}
\widehat{\Ab}^\top \!\Big( \Pb- \Pb\,\hat{\bb}\,\big(r+\hat{\bb}^\top \Pb\,\hat{\bb} \big)^{-1} \hat{\bb}^\top \Pb\Big)\,\widehat{\Ab} - \Pb + \Qb = \zerob
\end{equation}
for the exactly linearized system. This choice eventually leads to the subproblem of Example~\ref{exmp:quad} being a QCQP and the subproblems of Example~\ref{exmp:lin} being QPs. In Example~\ref{exmp:cos}, no special type of nonlinear program is obtained, but the costs are still chosen quadratic for simplicity and better comparison. Furthermore, it allows to choose 
$$
    \kappa(\xb) = -\big(r+\hat{\bb}^\top \Pb\,\hat{\bb} \big)^{-1} \hat{\bb}^\top \Pb\,\widehat{\Ab}\xb := \kappab_{\text{LQR}}^\top\xb,
$$
i.e., the control law of the linear quadratic regulator (LQR), and thus to represent exactly the cost of the unconstrained problem for an infinite horizon $N$ by the terminal cost $\varphi(\xb)$. Furthermore, we choose the terminal set
$$
\Tc \!= 
    \left\{\xb\in\R^n \left|\begin{pmatrix}
\Ib \\
\kappab_{\text{LQR}}
\end{pmatrix}\! \big(\widehat{\Ab} + \hat{\bb} \kappab_{\text{LQR}}^\top \big)^{k} \xb \in \Zc_1, \,\forall k \in \N\right.\right\}\!,
$$
which is known to inherit convexity from $\Zc_1$, to be efficiently computable under mild conditions \cite{Gilbert1991}, and to satisfy the axioms \eqref{eq:stability_axioms} for closed-loop stability.
Finally, the OCPs in every example are solved for a prediction horizon $N = 15$ as in \cite{NMPC2021Exact}, and the resulting feasible sets $\Fc$, terminal sets $\Tc$, as well as the optimal control law $u^\ast(\xb)$ and the associated optimal cost $V^\ast(\xb)$ are illustrated in the following.
\begin{figure}[H]
        \centering
        \includegraphics[trim=0.3cm 0cm 0.8cm 0.2cm,clip=true]{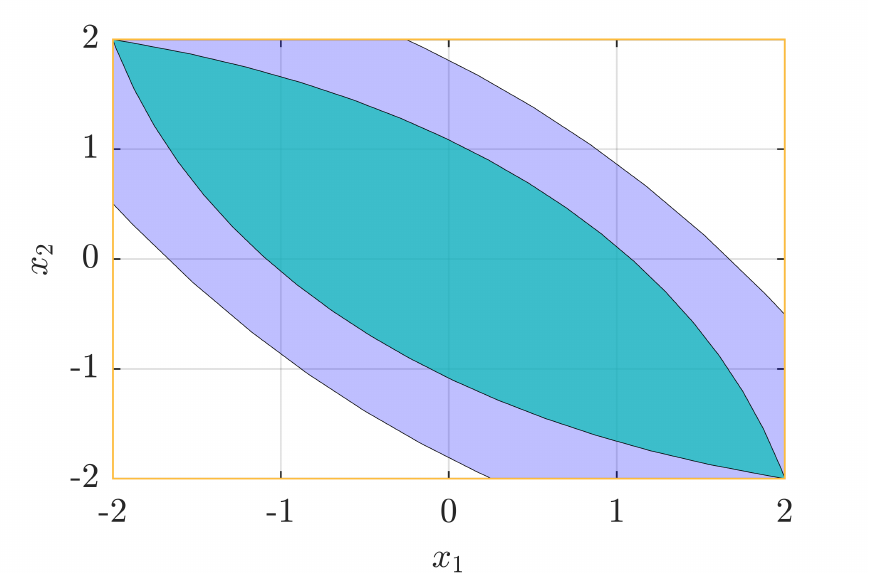}
        \vspace{-2mm}
        \caption{Illustration of the feasible set $\Fc$ (in blue) and the terminal set $\Tc$ (in cyan) for Example~\ref{exmp:quad}. The boundary of the set $\Xc$ is shown in orange.}%
        \label{fig:sets_quad}
        \centering
        \includegraphics[trim=6cm 10.7cm 6.4cm 10.8cm,clip=true]{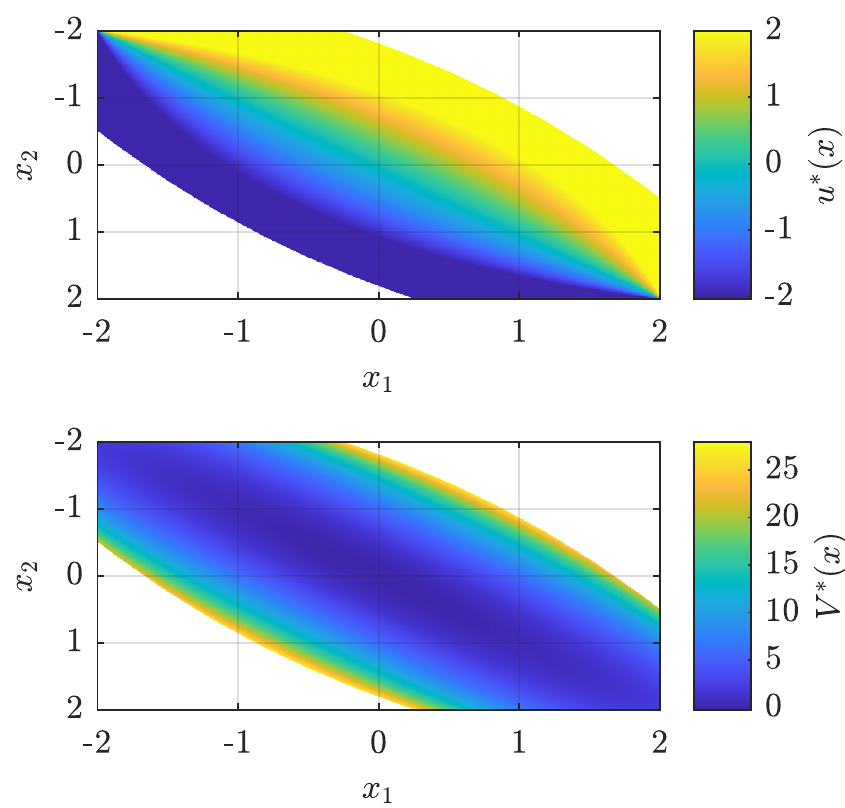}
        \vspace{-2mm}
        \caption{Illustration of the optimal control law $u^\ast(\xb)$ (top) and the optimal value function $V^\ast(\xb)$ (bottom) for Example~\ref{exmp:quad}.}
        \label{fig:law_cost_quad}
        \vspace{-2.5mm}
    \end{figure}

\begin{figure}[tp]
        \centering
            \includegraphics[trim=0.1cm 0cm 0.8cm 0.2cm,clip=true]{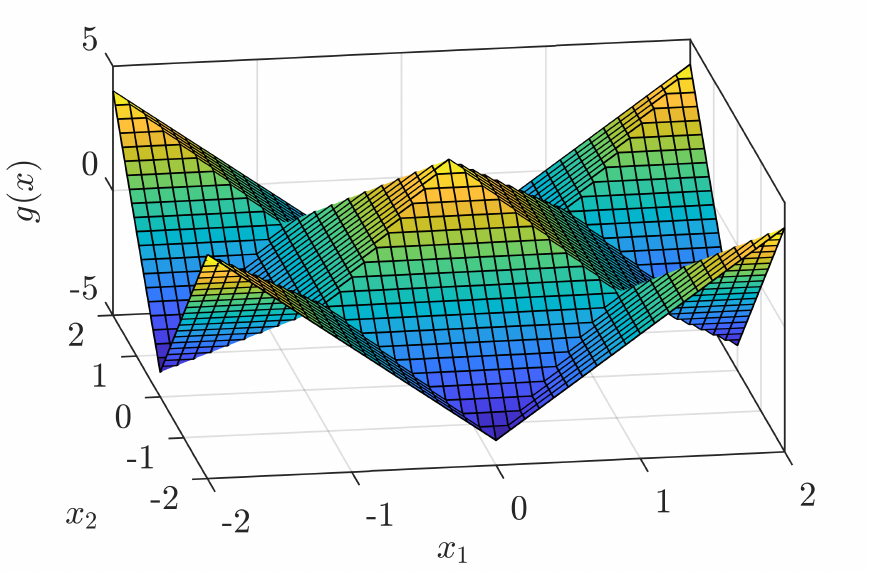}
            \vspace{-2mm}
            \caption{PWA function $g(\xb)$ considered in Example~\ref{exmp:lin}.}
            \label{fig:g_PWA}
        \centering
        \includegraphics[trim=0.3cm 0cm 0.8cm 0.2cm,clip=true]{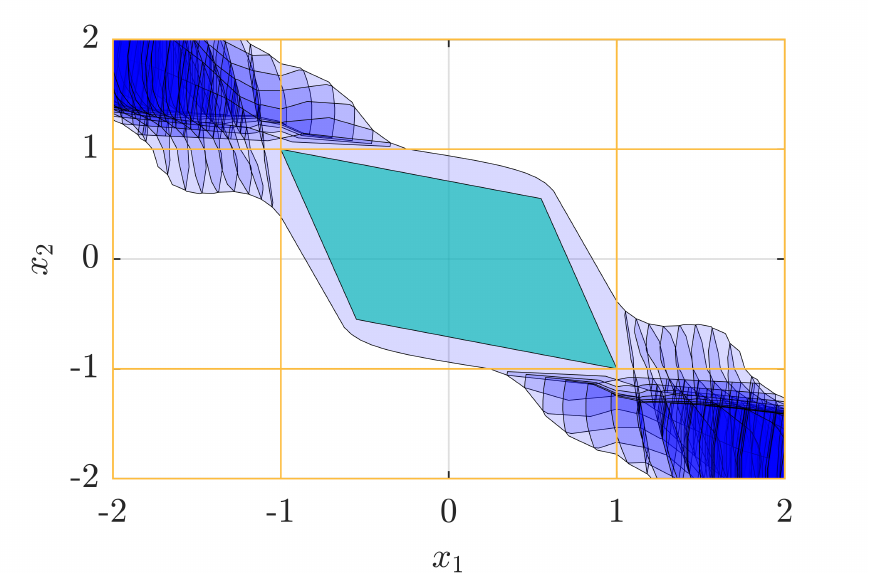}
        \vspace{-2mm}
        \caption{Illustration of the (partially overlapping) feasible sets $\Fc_j$ (in blue) and the terminal set $\Tc$ (in cyan) for Example~\ref{exmp:lin}. The boundary of the sets $\Xc_i$ is shown in orange.}
        \label{fig:sets_lin}
            
        \centering
        \includegraphics[trim=6cm 10.7cm 6.4cm 10.8cm,clip=true]{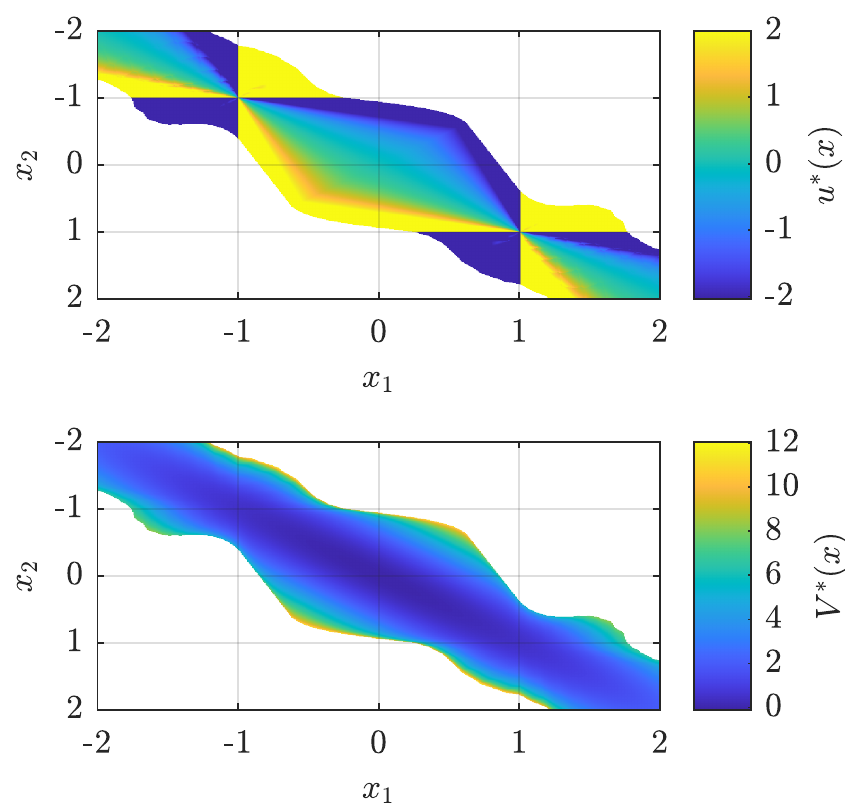}
        \vspace{-2mm}
        \caption{Illustration of the optimal control law $u^\ast(\xb)$ (top) and the optimal value function $V^\ast(\xb)$ (bottom) for Example~\ref{exmp:lin}.}
        \label{fig:law_cost_lin}
    \end{figure}

\begin{exmp}\label{exmp:quad}
    
    Consider the quadratic function 
    $$g(\xb) = \frac{3}{64}\xb_1^2-\frac{1}{8}\xb_1\xb_2+\frac{3}{64}\xb_2^2-2,$$
    which is convex and non-positive on the set $\Xc_1 = \Xc$. Hence, a decomposition into subsets is not required here. Since there is only a single convex subset, there also exists only $1^{15} = 1$ constraint scenario and thus only one convex subproblem, namely the original OCP \eqref{eq:MPC} itself. Because $g(\xb)$ is quadratic, the resulting inequality constraint functions in \eqref{eq:firstInequality} and \eqref{eq:secondInequality} are also quadratic functions of the optimization variables. Hence, with the chosen quadratic cost functions $\hat\ell(\xb,v)$ and $\varphi(\xb)$, the OCP takes the form of a QCQP and can thus be solved quite efficiently, especially since there is only a single constraint scenario to consider. A polytopic inner approximation of the feasible set $\Fc$ and the terminal set $\Tc$ is shown in Figure~\ref{fig:sets_quad}. 
    
    The resulting optimal control law $u^\ast(\xb)$ and the associated optimal cost $V^\ast(\xb)$ are illustrated in Figure~\ref{fig:law_cost_quad}. Both were obtained by sampling the constrained statespace $\Xc$ and implicitly solving the OCP. Note that, since $g(\xb) = 0$ for the two points $\xb = \begin{pmatrix}
        -2 & 2    
    \end{pmatrix}^\top$ and $\xb = \begin{pmatrix}
        2 & -2    
    \end{pmatrix}^\top$ that lie on the boundary of $\Xc$, the input becomes increasingly cheap towards these points, leading to a more aggressive optimal control $u^\ast(\xb)$. However, at these very points, where $\xb\in\Xc\setminus\Xc^\circ$, the optimal control law is set to $u^\ast = 0$ in accordance with Lemma~\ref{lem:equivalent_constraints}.
    More obvious cases of this behavior can be seen in the following examples, where $g(\xb) = 0$ not only occurs at isolated points on the boundary of $\Xc$.
\end{exmp}
\vspace{2mm}
\begin{exmp}\label{exmp:cos}

    Consider the sinusoidal function \vspace{1.5mm}
    $$
    g(\xb) = 4\cos\left(\frac{3\pi}{8}(\xb_1-\xb_2)\right) \vspace{1.5mm}
    $$
    which is concave and non-negative on the subset \vspace{1.5mm}
    $$
        \Xc_1 = \left\{\xb\in \Xc \left|-\frac{4}{3}\leq \xb_1-\xb_2 \leq \frac{4}{3}\right.\right\} \vspace{1.5mm}
    $$
    with $\zerob \in \interior{(\Xc_1)}$, as well as convex and non-positive on the subsets \vspace{1.5mm}
    \begin{align*}
        \Xc_2 &= \left\{\xb\in \Xc \left|-4\leq \xb_1-\xb_2 \leq -\frac{4}{3}\right.\right\} \quad \text{and} \\
        \Xc_3 &= \left\{\xb\in \Xc \left|\frac{4}{3}\leq \xb_1-\xb_2 \leq 4\right.\right\}. \vspace{1.5mm}
    \end{align*}
    Since we have $s = 3$ subsets, we obtain a total of $3^{15} \approx 14.3 \cdot 10^{6}$ convex subproblems, of which, however, only $31$ turn out to be feasible. The subsets $\Xc_i$ and a polytopic inner approximation of both the associated feasible sets $\Fc_j$ as well as the terminal set $\Tc$ are shown in Figure~\ref{fig:sets_cos}.
    
    Since all infeasible constraint scenarios are associated with empty feasible sets, the overall feasible set $\Fc$ of the NMPC problem is represented by the union of the illustrated sets. Remarkably, there exist feasible subproblems for constraint scenarios in which the state trajectory leaves $\Xc_1$ or $\Xc_3$ and enters $\Xc_2$, but not vice versa. The same one-way feature was observed in \cite{NMPC2021Exact} and is strongly associated with the input $u$, which becomes weaker as we have $g(\xb) = 0$ on the hyperplanes seperating $\Xc_i$, so that possible directions depend mainly on $A$. 
    
    The resulting optimal control law $u^\ast(\xb)$ and the associated optimal cost $V^\ast(\xb)$ are illustrated in Figure~\ref{fig:law_cost_cos}. Again, both were obtained by sampling the constrained statespace $\Xc$ and implicitly solving the OCP. Note that the optimal control $u^\ast(\xb)$ becomes increasingly aggressive near the hyperplanes separating $\Xc_1$, $\Xc_2$ and $\Xc_3$, where we have $g(\xb) = 0$ by construction of the subsets $\Xc_i$. Furthermore, because $g(\xb)$ changes sign exactly at these two hyperplanes, the optimal control $u^\ast(\xb)$ jumps from its maximum $\overline u$ to its minimum $\underline u$ and vice versa, with $u^\ast(\xb) = 0$ for $g(\xb) = 0$ as noted in Lemma~\ref{lem:equivalent_constraints}. We also note that both $u^\ast(\xb)$ and $V^\ast(\xb)$ may be discontinuous not only at these hyperplanes but also at any part of $\Xc$ with overlapping feasible sets $\Fc_j$. This can be intuitively explained by the fact that for a given state $\xb$ the optimal value $V^\ast(\xb)$ can be obtained for more than a single constraint scenario at a time, which is consistent with the fact that the optimal control sequence of a nonconvex OCP is not guaranteed to be a singleton. A more formal argument can be made by considering the OCP as a  multiparametric nonlinear program, for which continuity conditions are stated in, e.g., \cite{borrelli_bemporad_morari_2017}.
    
    \begin{figure}[tp] \vspace{-0.5mm}
        \centering
        \includegraphics[trim=0.3cm 0cm 0.8cm 0.2cm,clip=true]{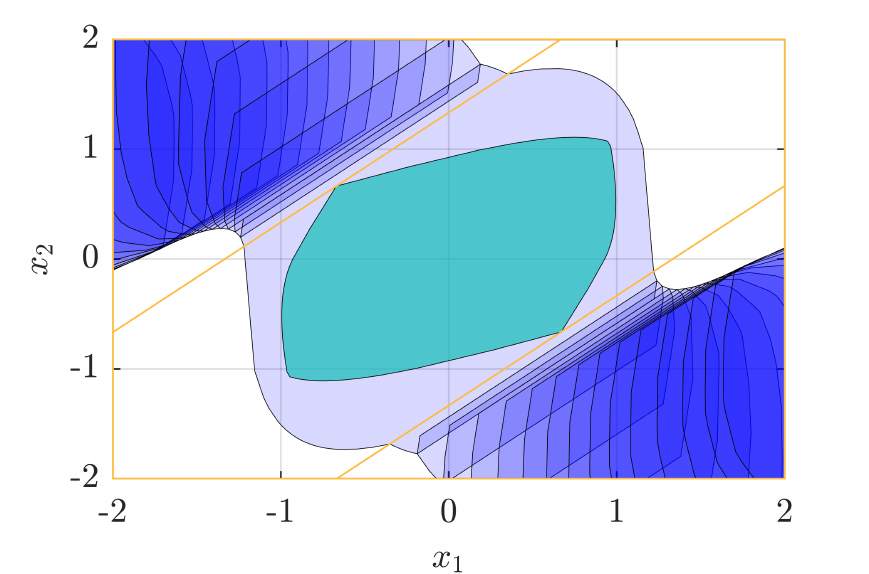}
        \vspace{-2mm}
        \caption{Illustration of the (partially overlapping) feasible sets $\Fc_j$ (in blue) and the terminal set $\Tc$ (in cyan) for Example~\ref{exmp:cos}. The boundary of the sets $\Xc_i$ is shown in orange.}
        \label{fig:sets_cos}\vspace{4mm}
        \centering
        \includegraphics[trim=6cm 10.7cm 6.4cm 10.8cm,clip=true]{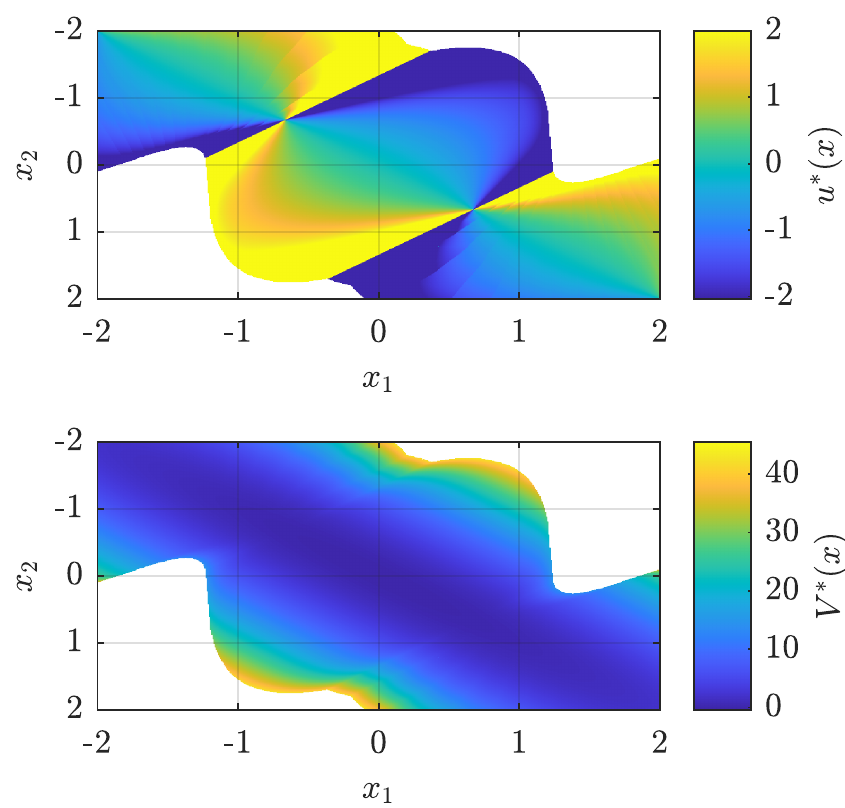}
        \vspace{-2mm}
        \caption{Illustration of the optimal control law $u^\ast(\xb)$ (top) and the optimal value function $V^\ast(\xb)$ (bottom) for Example~\ref{exmp:cos}.}
        \label{fig:law_cost_cos}\vspace{0.5mm}
    \end{figure}
    \end{exmp}
    \begin{exmp}\label{exmp:lin}
        Consider a continuous PWA function
        $$
            g(\xb) = \wb_{\mu}^\top\xb+d_{\mu} \quad \xb \in \Sc_{\mu}(\xb) 
        $$
        defined on polytopic partitions $\Sc_{\mu}(\xb)$ of $\Xc$. Since affine functions are both convex and concave, it is easy to construct polytopic convex subsets $\Xc_i$ such that Assumption~\ref{assum:characterization} holds. Note that, similar to \cite{NMPC2021Exact}, it can be shown that the resulting $\Zc_i$ are not only convex but polyhedral, and therefore so is the terminal set $\Tc$ \cite{Gilbert1991}. It follows that the resulting convex subproblems have linear inequality constraints as well as quadratic costs, and thus take the form of QPs.

        For this example, we consider the PWA function shown in Figure~\ref{fig:g_PWA} which may be defined on a total of $12$ partitions $\Sc_{\mu}(\xb)$, but for brevity the definitions are not explicitly written down here. Note that the subsets $\Xc_i$ need not correspond to these partitions $\Sc_{\mu}(\xb)$ of the PWA definition, but instead we choose the largest possible subsets for which Assumption~\ref{assum:characterization} is satisfied. This gives $s = 9$ subsets and thus a total number of $9^{15} \approx 2.1 \cdot 10^{14}$ convex subproblems, which, however, result in only $217$ feasible subproblems after recursive inspection. The chosen subsets $\Xc_i$ and both the feasible sets $\Fc_j$ as well as the terminal set $\Tc$, which were computed exactly here since only linear inequality constraints occur, are shown in Figure~ \ref{fig:sets_lin} and, again, the overall feasible set $\Fc$ is represented by the union of the illustrated $\Fc_j$. The same one-way feature as in Example~\ref{exmp:cos} is observed here, which helps to explain the small number of feasible constraint scenarios compared to their total number.
        
        Since in this example the underlying subproblems for each constraint scenario correspond to linear MPC problems, $V^\ast(\xb)$ and $v^\ast(\xb)$ were obtained by evaluating explicit solutions as described in \cite{Bemporad2000}. The resulting optimal control law $u^\ast(\xb)$ and the associated optimal cost $V^\ast(\xb)$ are illustrated in Figure~\ref{fig:law_cost_cos}. Again, we observe discontinuity and increasing aggressiveness of the optimal control $u^\ast(\xb)$ for $g(\xb)$ approaching zero.
  \end{exmp}
  
\section{Summary and Outlook}\label{sec:summary}
We showed that a special class of (nonconvex) NMPC problems can be reformulated as the solution of a finite number of convex subproblems. These results can be obtained for a special subclass of input-affine nonlinear discrete-time systems \eqref{eq:specialDynamics} satisfying Assumption~\ref{assum:characterization} by utilizing exact (input-state) linearization. While exact linearization alone does not usually lead to a convex OCP, we have shown how the cost functions and a decomposition of the constraints can be chosen to allow a convex reformulation. We applied these results to three numerical examples, where we also show that the resulting subproblems can take certain standard forms of mathematical programs depending on the structure of the system itself.

While the shown results generalize previous findings of \cite{NMPC2021Exact} to a much broader class of systems, the applicability is still very much limited due to the restrictive Assumption~\ref{assum:characterization}. Future research on this topic should aim to generalize it even further or find similar kinds of convex reformulations for other classes of systems. A natural next step would be, e.g., the treatment of multi-input systems.


\end{document}